\documentstyle[sprocl]{article}

\def\be{\begin{equation}}
\def\ee{\end{equation}}
\def\bea{\begin{eqnarray}}
\def\eea{\end{eqnarray}}

\begin{document}
\title{SIMULATIONS OF CORE COLLAPSE SUPERNOVAE\\ IN ONE AND TWO DIMENSIONS\\ USING 
MULTIGROUP NEUTRINO TRANSPORT}
\author{A. MEZZACAPPA}
\address{ORNL, Physics Division, Building 6010, MS 6354, P.O. Box 2008\\
Oak Ridge, TN 37831-6354}
\maketitle\abstracts{
In one dimension, we present results from comparisons of stationary 
state multigroup flux-limited diffusion and Boltzmann neutrino transport, 
focusing on quantities central to the postbounce shock reheating. In 
two dimensions, we present results from simulations that couple 
one-dimensional multigroup flux-limited diffusion to two-dimensional 
({\small PPM}) hydrodynamics. 
}

\section{Introduction}
Current supernova modeling revolves around the idea that the
stalled supernova shock is reenergized by absorption of 
electron neutrinos and antineutrinos on the shock-liberated
nucleons behind it.\cite{bw85} Key to this process is the
neutrino transport in the critical semitransparent region 
encompassing the neutrinospheres. Potentially aiding this 
process is convection below the neutrinospheres and the 
shock,\cite{bhf95,hbhfc94,jm96,wm93} although the verdict 
has been mixed.\cite{bhf95,jm96,m96a,m96b,mwm93}

In semitransparent regions, neutrino transport approximations 
such as multigroup flux-limited diffusion ({\small MGFLD}) 
break down. Moreover, neutrino shock reheating/revival depends 
sensitively on the emergent neutrinosphere luminosities and 
spectra, and on the neutrino inverse flux factors between the 
neutrinospheres and the shock. At the very least, computation 
of these quantities by approximate transport methods have to 
be checked against exact methods, i.e., against Boltzmann 
neutrino transport. 

We present results from one- and two-dimensional supernova
simulations. 
In one dimension, beginning from postbounce slices obtained 
from Bruenn's {\small MGFLD} simulations, which are subsequently
thermally and hydrodynamically frozen, we compute stationary 
state neutrino distributions with {\small MGFLD} and Boltzmann
neutrino transport. From these, we compute and compare the
neutrino luminosities, {\small RMS} energies, and inverse
flux factors, and the net neutrino heating rate, between
the neutrinospheres and the shock.
In two dimensions, beginning with the same postbounce profiles,
we couple one-dimensional {\small MGFLD} to two-dimensional
{\small PPM} hydrodynamics to investigate ``prompt'' convection
below the neutrinospheres and ``neutrino-driven'' convection 
below the shock. 
At the expense of dimensionality, we implement neutrino transport 
that is multigroup and that computes with sufficient realism 
transport through opaque, semitransparent, and transparent 
regions. 

\section{Prompt Convection}
Without neutrino transport, prompt convection 
develops and dissipates in $\sim$15 ms in both our 15 and
25 M$_{\odot}$ models. When transport is included, there 
is no significant convective transport of entropy or 
leptons in either model. Neutrino transport {\it locally}
equilibrates a convecting fluid element with its surroundings,
reducing the convection growth rate and asymptotic convection
velocities by factors of 4--250 between $\rho=10^{11-12}$ 
g/cm$^{3}$, respectively.\cite{m96a} Our transport neglects the finite
time it takes for neutrinos to traverse a convecting fluid
element; consequently, our equilibration rates are too rapid,
and transport's effect on prompt convection is overestimated.
Nonetheless, it is difficult to see how prompt convection will
lead to significant convective transport when these 
effects are taken into account.

\section{Neutrino-Driven Convection}
In our 15 M$_{\odot}$ model, large-scale semiturbulent convection 
is evident below the shock and is most vigorous at $t=225$ ms
after bounce (we began our run at 106 ms after bounce when a 
well-developed gain region was present.) 
At this time, the maximum angle-averaged entropy is 13.5, and the 
angle-averaged radial convection velocities exceed $10^{9}$ cm/sec,
becoming supersonic just below the shock. Despite this, at $t=506$ 
ms after bounce our shock has receded, the convection below it 
has become more turbulent, and there is no evidence of an
explosion or of a developing explosion.\cite{m96b} 

Different outcomes obtained by the various groups are mainly 
attributable to differences in the neutrino transport approximations, 
which determine the postbounce initial models and the neutrino 
luminosities, {\small RMS} energies, and inverse flux factors, which
in turn define the postshock neutrino heating rates. 
Most notable are dramatic differences in the {\small RMS}
energies.\cite{m96b} For example, for $\eta_{\nu_{\rm e}}=2$ 
and T$_{\nu_{\rm e}}=$T$(\tau_{\nu_{\rm e}}=2/3)$,
which is implemented by Burrows et al.
(1995),\cite{bhf95} the electron neutrino {\small
RMS} energy and matter temperature at the neutrinosphere are related by
$<E_{\nu_{\rm e}}^{2}>^{1/2} = 3.6$T, whereas
our {\small MGFLD} calculations give 
$<E_{\nu_{\rm e}}^{2}>^{1/2} = 3.0$T. 
Because the neutrino heating
rates depend on the square of the {\small RMS} energies,
the Burrows et al. rates would effectively be 40--50\% higher. 
In the Herant et al. (1994)\cite{hbhfc94} calculations, at 100 ms 
after bounce $<E_{\nu_{\rm e}}>\sim$ 13 to 14 MeV and 
$<E_{\bar{\nu}_{\rm e}}>\sim$ 20 MeV, whereas 
we obtain the significantly lower values, 10 and 
13 MeV, respectively. 

\section{Boltzmann Neutrino Transport}
Although differences in the {\small MGFLD} and Boltzmann 
transport luminosities and {\small RMS} energies between 
the neutrinospheres and the shock were seen, the inverse 
flux factors differed most. The fractional difference 
relative to {\small MGFLD} rose to $\sim$ 20--25\% just 
below the shock. 
The net heating rate showed an even more pronounced
difference, with the Boltzmann rate being 
2--3 times higher just above the gain radius.\cite{m97} 
(Small differences in the heating rates lead to large
differences in the net heating rate near the gain
radii, where heating and cooling balance.)
These results have two ramifications: (1) It may be
possible to obtain explosions in spherical symmetry
in the absence of convection.\linebreak (2) Improved 
transport in two and three dimensions will
give rise to more neutrino heating, and may give rise 
to more vigorous neutrino-driven convection. 
Simulations in both spherical and axi- symmetry with 
one-dimensional Boltzmann transport are planned. We 
are also developing realistic multidimensional, multigroup 
transport.

\section*{Acknowledgments} 
AM is supported at the Oak Ridge National Laboratory, which is 
managed by Lockheed Martin Energy Research Corporation under DOE 
contract DE-AC05-96OR22464. This work was also supported at the 
University of Tennessee under DOE contract DE-FG05-93ER40770.

\end{document}